\documentclass[twocolumn,prl]{revtex4}
\usepackage{graphicx}
\begin{document}

\title{From deterministic dynamics to kinetic phenomena}

\author{S. Denisov, A. Filippov, J. Klafter, M. Urbakh}
\address{\it School of Chemistry, Raymond and Beverley Sackler Faculty of Exact Sciences,
Tel-Aviv University, \it
Tel-Aviv 69978, Israel}
%%\date{8th March 2002}

\date{\today}

\begin{abstract}
    We investigate  a one-dimenisonal Hamiltonian system that
describes a system of particles interacting through short-range
repulsive potentials. Depending on the particle mean energy,
$\epsilon$, the system demonstrates a spectrum of kinetic regimes,
characterized by their transport properties ranging from ballistic
motion to localized oscillations through anomalous diffusion
regimes.  We etsablish relationships between the observed kinetic
regimes and the "thermodynamic" states of the system.  The nature
of heat conduction in the proposed model is discussed.
\end{abstract}
\maketitle

The question of how statistical laws emerge from microscopic
dynamics has been a subject of interest for a long time [1].
Studies of relatively simple dynamical systems provide a link
between deterministic dynamics and diffusion phenomena [2,3]. In
particular, a number of recent dynamical models aim at
understanding the necessary and sufficient conditions for a system
to obey the Fourier heat conduction law [4]. These examples cover
only a part of general problem of how kinetic and thermodynamic
properties emerge from dynamics at the atomic scale.

In this Letter we propose a many-particle Hamiltonian model which
exhibits a wide range of mass and energy transport regimes, and
allows to establish relationships between microscopic properties
and thermodynamical and kinetic phenomena. We demonstrate that
kinetic properties of the system are determined by one controlling
parameter, which is a mean energy per a particle, $\epsilon$.
Varying $\epsilon$ one can cover the whole spectrum of diffusion
regimes, from ballistic motion to "frozen" states through
anomalous diffusion regimes. We show that the observed kinetic
regimes are strongly related to  the "thermodynamic" states of the
system which change from a "solid" to "gas" phase as the parameter
$\epsilon$ increases.

The model describes a system of $N$ classical particles each
characterized by  coordinate $x_{i}$  and the conjugate momentum
$p_{i}$. The particles interact through a repulsive short range
potential  according to the following Hamiltonian
\begin{eqnarray} \label{1}
H(x_{i}, p_{i}) =\sum_{i=1}^{N}
\frac{p_{i}^{2}}{2}+\frac{A}{2}\sum_{i,j=1}^{N}
e^{(-(x_{i}-x_{j})^2/\sigma^{2})},
\end{eqnarray}
where $A$ is a strength of inter-particle interaction and $\sigma$
is a width of repulsive core. The particles are located on an
interval of length $L$, and  periodic boundary conditions are
applied. The mean energy and density of particles are defined as
$\varepsilon =E/N$ and $\varrho=N/L$, respectively.

In the  high energy limit, $\varepsilon /A \gg 1$, the system in
Eq.(1) behaves as a gas of freely flying particles, slightly
perturbed by weak interactions. In the opposite  low energy limit,
$\varepsilon /A \ll 1$, the system forms a crystal lattice with a
constant $a_{0}=1/\varrho$. In this solid-like state each particle
oscillates at the minimum of potential well formed by its
neighbors.  In the gas phase the system is uniform while in the
solid state it has a discrete translational symmetry with a
lattice constant $a_{0}$. According to Kolmogorov - Arnol'd -
Moser theorem, in both limits, the system exhibits an ordered
dynamical behavior in a region of positive measure in  phase space
[6]. Due to a difference in symmetries of the ordered states one
can expect that a transition between these two states with the
change of $\epsilon$ will occurs through a mixed disordered state.

To study the dynamics of the system in Eq.(1) we use the following
algorithm: At the beginning we put all particles at an equal
distance $a_{0}$ from each other and give them a kinetic energy
according to the Maxwell distribution. After that we rescale
velocities of the particles in order to get a total energy equal
to $N \cdot \epsilon$. Then, for each time step, we integrate the
corresponding dynamical equations using a central difference
symplectic scheme [5]. To extract an information on the structure
and dynamics we introduce also a sorted array of particles, $\{
x^{sort}_{j}(t)\}$, where particles are renumbered according their
actual instantaneous position, $j=sort(i)$.

In order to illustrate a nature of excitations in the system we
show in Fig.1a a time evolution of energy distribution initiated
by a local perturbation  at time $t=0$ that had a form of a kicked
group of a few central particles. Here particles with high and low
instantaneous velocities are displayed by light and dark colors
respectively.  The light regions are organized into lines which
correspond to excitations propagating along the system. Three
types of excitations with different group velocities, which are
given by an inverse slope of the lines, can be distinguished:
flying particles, low-energy phonons and nonlinear solitonic
excitations which are intermediate in energy between them. The
first dynamical mode, flying particles, dominates in the gas
state, $\epsilon/A \gg 1$. Nonlinear excitations and phonons play
the main role in the opposite solid-state limit, $\epsilon /A \ll
1$, (Fig.1c). An energy exchange between different  modes is most
effective in a "liquid-like" state, that corresponds to $\epsilon
/A \approx 1$. Interactions between modes can lead, for example,
to a transformation of flying particle into solitonic excitation
(see a circle in Fig.1a) or to "burning" of flying particle from a
sea of phonons and nonlinear excitations (see a circle in Fig.1b)

In order to describe quantitatively a structure of different
"thermodynamic" states arising in the system, we introduce a
reduced distance for a sorted set of particles,
$\xi_{j}=(x^{sort}_{j}(t)-x^{sort}_{j-1}(t)-a_{0})$, and calculate
the probability density function (pdf) $\Phi(\xi)$, $-a_{0} < \xi
< L/2$ (see Fig.2). In the low energy limit the pdf $\Phi(\xi)$ is
governed by a peak near the point $\xi=0$, which corresponds to a
crystal lattice (a right inset in Fig.2). In the gas limit the
density of states scales as (see left inset in Fig.2)
\begin{eqnarray} \label{1}
\Phi(\xi) \sim e^{-\chi (\xi-a_{0})}.
\end{eqnarray}

\begin{figure}[t]
\includegraphics[width=0.9\linewidth,angle=0]{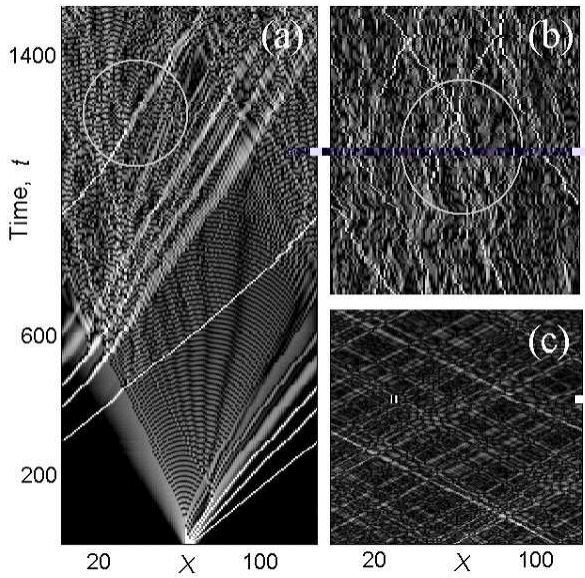}
\caption{ (a) Time evolution of the energy distribution initiated
by kicking of few central particles, for the system in Eq. (1)
($N=128$, $\epsilon / A=0.25$).  The circle indicates a collapse
of flying particle into a nonlinear excitation. (b) The same
system after time $t=5000$. The circle indicates a birth of two
flying particles from the excitation sea. (c) The set of
propagating solitons for the system in Eq. (1) for $\epsilon/A=1$.
\label{fig1}}
\end{figure}

It is natural to assume that in the intermediate, liquid state the
pdf, $\Phi(\xi)$, can be presented in the form
\begin{eqnarray} \label{1}
\Phi(\xi)=\Phi_{solid}(\xi)+\Phi_{gas}(\xi)
\end{eqnarray}
where $\Phi_{solid}(\xi)$ (a dashed line in Fig.2)  and
$\Phi_{gas}(\xi)$ (a thin line in Fig.2) describe a solid-state
and  gas contributions correspondingly. The solid and gas
contributions can be isolated from the net pdf by fitting its
asymptotics to the exponential distribution, Eq.(2), at negative
and large positive $\xi$. Then Eq.(3) allows to introduce the
order parameter, which presents a fraction of solid  phase
\begin{eqnarray} \label{1}
\phi=\frac{\int_{-a_{0}}^{L/2}\Phi_{solid}(\xi)d\xi}{\int_{-a_{0}}^{L/2}\Phi(\xi)d\xi}.
\end{eqnarray}

Fig.3a shows a dependence of the order parameter $\phi$ on the
energy density $\varepsilon$. It exhibits a phase transition from
the gas to solid state as $\epsilon/A$ decreases. The plateau at
$\epsilon/A \approx 1$ indicates on the presence of the third,
mixed state that can be associated with a liquid-like phase. Below
we consider regimes of mass and energy transfer which correspond
to the different thermodynamic states.

{\it Flying particles.}  Particles can fly only when their
velocities are higher than a threshold value, $v_{fl} \approx
\sqrt{2U_{max}}$, which is determined  by the height of the
effective potential created by neighboring particles. The latter
can be estimated as a barrier in a periodic lattice, formed by
particles in the solid state.

Each particle dynamics can be characterized by the mean squared
displacement $\langle x^{2}(t) \rangle$, which in the long-time
limit, follows
\begin{eqnarray} \label{1}
\langle x^{2}(t) \rangle \sim t^{\alpha}
\end{eqnarray}
where $\alpha=1$ for normal diffusion. All processes with $\alpha
\neq 1$ are known as anomalous diffusion [7,8], being subdiffusion
for $0<\alpha<1$ and superdiffusion  for $1 < \alpha < 2$.

\begin{figure}[t]
\includegraphics[width=1.\linewidth,angle=0]{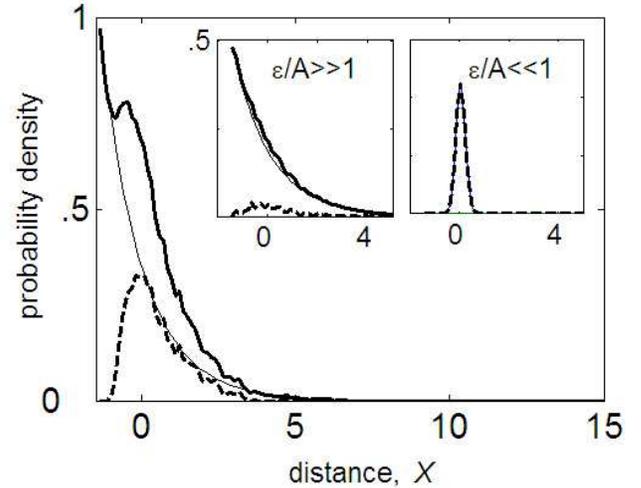}
\caption{  The pdf $\Phi(\xi)$ calculated  for the liquid state,
$N=64$, $A=1$, $\epsilon=0.25$. The thin line shows fitting to
$\Phi_{gas}(\xi)$ in Eq.(2). Dashed line corresponds to
$\Phi_{solid}(\xi)=1-\Phi_{gas}(\xi)$. Left inset shows pdf
$\Phi(\xi)$ in gas state ($\epsilon/A=0.1$) and the rigth inset
corresponds to solid state ($\epsilon/A=0.1$).
 \label{fig2}}
\end{figure}

The numerical results  presented in Fig.3a clearly show three
different dynamical regimes: (1) Absence of diffusion for
$\epsilon /A \ll 1$ (solid phase). In this case particles perform
small oscillations around the minima of the potential, and as a
result $\alpha=0$. (2) A ballistic regime with $\alpha=2$ for
$\epsilon /A \gg 1$ (gas phase).  (3) Normal diffusion,
$\alpha=1$, in the liquid-like state where $\epsilon/A \sim 1$. It
should be noted that the latter regime exists for a finite
interval of $\epsilon$. For the values of $\epsilon$ corresponding
to gas-liquid and liquid-solid transitions we have found regimes
of anomalous diffusion, namely a superdiffusion in the first case
and subdiffusion the second case (Fig.3)[9].

In the case of superdiffusion  each particle performs both flights
and trapped oscillations, randomly switching between these two
dynamical modes (see Fig.1a and b). We have obtained the histogram
for a single flight time $\Psi(t)$, accumulating the lifetime and
length during the interval where the sign of the particle velocity
is fixed.  It shows an asymptotic power law decay (Fig.3b),
$\Psi(t) \sim t^{-\gamma}$, with the exponent $\gamma \approx 2.5$
for the gas phase ($\epsilon/A=10$) that corresponds to
superdiffusion and $\gamma \approx 3$ for the high energy edge of
the liquid phase, at $\epsilon/A=2$, that corresponds to a
transition to normal diffusion regime. The related pdf has the
same power behaviour. We note that the flying particles do not
have a constant velocity.

\begin{figure}[t]
\includegraphics[width=0.95\linewidth,angle=0]{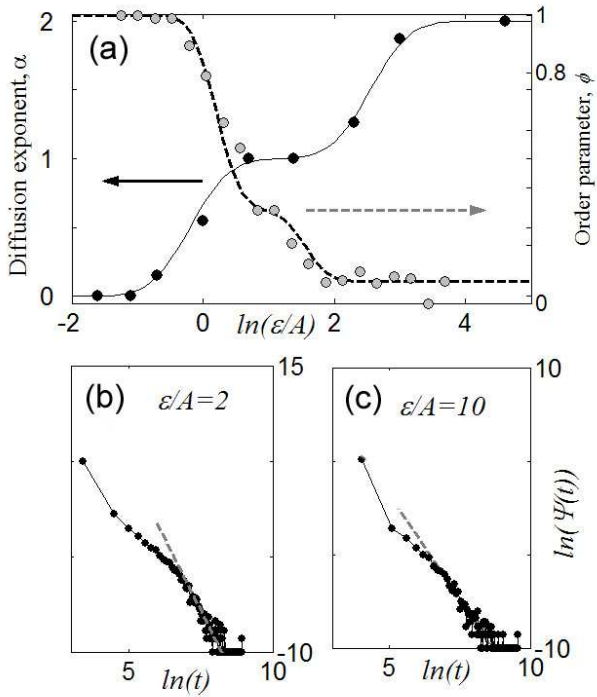}
\caption{ (a) The dependence of the order parameter, Eq. (4),
$\phi$ (black circles) and  the diffusion exponent $\alpha$, Eq.
(5) (grey circles),  obtained by fitting of msd, $\langle x^{2}(t)
\rangle$, for a calculation time $t=10^4$; The histograms for a
duration of a single flight, $\Psi(t)$, for (b) $\epsilon/A=0.2$
and (c) $\epsilon/A=1.2$; $N=512$.
 \label{fig1}}
\end{figure}

The subdiffusion, which has been found in the transition region
between solid and liquid states, corresponds to short random jumps
of the particles interrupted by long trapping events.  At low
energies the probability of strong fluctuations which can result
in a particle escape from the potential well becomes very low.
This leads to anomalously long trapping of particles.  The
particle motion becomes subdiffusive when the pdf of trapping
times has a power law asymptotic, $\Psi(t) \sim t^{-1-\beta}$ with
$\beta < 1$ [8]. In this case the diffusion exponent $\alpha$ in
Eq. (5) equals $\beta$, $\alpha=\beta < 1$. The subdiffusive
dynamics can be quantitatively described using a continuous-time
random walk (CTRW) approach [8].

{\it Solitary excitations and phonons.} In the low-energy regime
the system tends to build a crystal-type structure. Local energy
fluctuations can create here only relatively smooth nonlinear
excitations as well as small-amplitude phonons. In the phonon
state particles are mainly localized in the middle point between
two nearest neighbors, and the variable $ \xi_{j}$ introduced
earlier can be used as a small parameter, $\xi_{j}/a_{0} \ll 1$
[11]. Lianerization of the equations of motion for the Hamiltonian
in Eq.(1) gives the elastic constant
$C=\frac{2A}{\sigma^{2}}(1-\frac{2}{(\sigma\varrho)^{2}})e^{-\frac{1}{(\sigma\varrho)^{2}}}$,
which contains all information on the phonon spectrum [11].  This
approximation is not valid for nonlinear excitations, where
particles can strongly deviate from their equilibrium positions.
However, in the case of smooth excitations the displacements of
particles with respect to the corresponding middle points change
slowly with  particle number $j$ in the sorted array
$x^{sort}_{j}$, and the difference $y_{j}=( x^{sort}_{j+1} +
x^{sort}_{j-1})/2- x^{sort}_{j}$ can be considered as a smooth
variable $y=y(x,t)$. One can rewrite the Hamiltonian in Eq.(1) in
terms of the continuous variable $y(x,t)$ and prove that the
solitonic nonlinear excitations propagating with a constant
velocity, $y=y(x-vt)$, indeed exist in the system.

 We have found that such solutions
exist only for particle densities which are below a critical
density,  $\varrho_{crit} =1/a_{0crit}= \sqrt{2}/\sigma$. For the
densities higher than the critical one the width of the
excitations reduces to a single particle scale and the smooth
nonlinear excitations  cannot exist.  Under this condition the
liquid state (the plateau in Fig.3a) almost disappears and a
transition between solid and gas phases goes directly through a
"sublimation" process.  Numerical studies of the nonlinear
solutions show that they have properties of "quasi-particles".
They propagate in both directions practically without a loss of
energy (see Fig.1c) and collide with each other another.
Interactions between different types of excitations produce an
intensive energy exchange in the liquid state that leads to the
most efficient thermalization of the system in this state.

\begin{figure}[t]
\includegraphics[width=0.8\linewidth,angle=0]{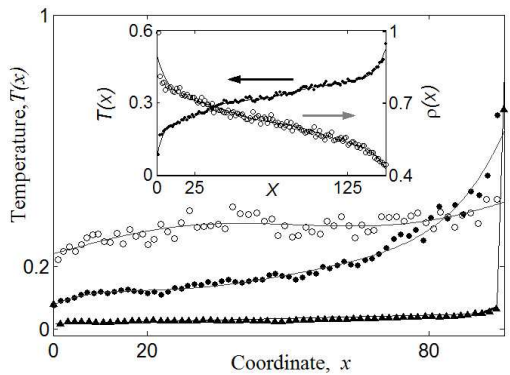}
\caption{ Local temperature profiles $T_{i}$ for $\epsilon/A=10$
(open circles), $\epsilon /A=1$ (filled circles), and
$\epsilon/A=0.1$ (triangles); $N=100$, $T_{l}=2$, $T_{r}=0.1$.
Inset shows local temperature profile and local density of
particle, calculated for $T_{l}=0.1, T_{r}=2$, $N=128$, $\epsilon
/A =0.25$. Lines obtained by averaging over $100$ realizations
after a transient time $t=10^{4}$.
 \label{fig1}}
\end{figure}

{\it Heat conduction.} The proposed system provides  a possibility
to simulate kinetic properties of classical low-dimension systems
[4]. As an example we consider a problem of heat conductivity. For
this purpose the system in Eq. (1)  is coupled to the heat
reservoirs placed at the walls $x=0$ and $x=L$. The temperatures
at the left and the right walls are given by $T_{l}$ and $T_{r}$
$(T_{l}>T_{r})$, respectively. When a particle collides with a
wall at temperature $T_{l,r}$, it is reflected back with a
velocity chosen from the distribution
$f(v)=(|v|/T_{l,r})exp(-mv^{2}/T_{l,r})$.  To calculate a
temperature profile we evaluate time averages as follows: we
divide the interval $L$ into a set of equal cells, ${X_{s}}$,
$s=1$, $N-1$, with a length $\Delta X=L/N$. In Fig.4 we show the
local temperature distribution $T(x)=\langle v^{2}_{s} \rangle/2$
found for different thermodynamic phases of the system in Eq. (1).
Here $\langle v_{s}^{2} \rangle/2$ is an average kinetic energy of
the $s$-cell and $\langle ~ \rangle$ denotes a time averaging. $s$
denotes the location along the interval and corresponds in a
continuum description to $x$. In the gas limit the temperature
profile exhibits a wide plateau which is typical for a ballistic
mechanism of conductivity (Fig.4, open circles). In this case the
heat transfer is determined by flying particles. In the opposite,
solid state, limit we also find ballistic conductivity which is
determined by propagating quasi-particles: solitonic excitations
and phonons (Fig.4, squares)[4]. In the liquid state the $T(x)$
has a strong nonlinear profile with a substantial nonzero slope
that is a result of intensive scattering processes and energy
exchange between all dynamical modes (Fig.4, filled circles). The
local density of particles, $\rho(x)=\langle n_{s}\rangle/N$,
where $n_{s}$ is the average number of particles per cell, also
demonstrates a nonlinear profile, which is complimentary to the
temperature profile.  The observed variation of $\rho(x)$ can be
explained by an accumulation of slow particles at the cold end and
by fast escape of flying particles from the hot end. A relation
between temperature and concentration gradients, which is usually
assumed for a macroscopic system close to equilibrium, naturally
emerges from the proposed Hamiltonian model.

Our calculations show that the proposed system can be used as a
model for {\it thermo-engine} that transforms  heat into a
directed motion. For this purpose the ensemble of particles should
be coupled to a third body, i. e. a cargo. Slow particles moving
from the cold end affect the cargo stronger than fast particles
that move from the hot end. This asymmetry leads to a directed
motion of the cargo. Another example of transformation of thermal
energy into a directed motion is a production of {\it electric
current} due to  temperature differences of thermostats. To
simulate this phenomenon we have considered two kinds of particles
with different masses and opposite charges.  When the difference
in the masses of the particles is large enough the light ones can
fly ballistically and transfer current while the heavy ones remain
immobile. A detailed consideration of phenomena related to a
directed motion will be published elsewhere.

In summary we have introduced a simple dynamical model  that
establishes the relationships between microscopic properties of
Hamiltonian systems and "macroscopic" thermodynamical and kinetic
phenomena. Depending on the energy the model exhibits three
well-defined states and a wide spectrum of kinetic phenomena. An
interesting and still unresolved question concerns a scaling of
Fourier coefficient for the thermoconductivity, $k\sim L^{\beta}$,
with a length of the system, $L$, [4].   The solution of this
problem requires much longer simulations than that reported in
this Letter. On the base of our calculations we expect that normal
heat conduction, $k \sim L^{0}$, should be observed in the liquid
state.

Financial support for this work by grants from the Israel Science
foundation (Grant No 573/00) and BSF is gratefully acknowledged.


\begin{thebibliography}{99}
\bibitem{9}
It should be noted that phases disscussed above present long-lived
out-of-equlibrium states. After a crossover time $t_{c} \sim
10^{5}$, the exponent changes to  $\alpha=1$ (normal diffusion).
We have found that the transient time increases with $N$ and the
out of equlibriium states become stationary solutions for  $N
\rightarrow \infty$. The link between relaxation to equilibrium
and anomalous diffusive motion in classical $N$ - particles
Hamiltonian systems with energy conservation have been discussed
in Ref. [10].
\end{thebibliography}
\end{document}